\documentclass[prb,twocolumn,showpacs,superscriptaddress]{revtex4}

\usepackage{graphicx,amsmath}
\usepackage{bm}

\newcommand{\ij}{i\kern -0.08em j}
\newcommand{\half}{{\textstyle\frac{1}{2}}}
\def\hn{\mskip-0.5\thinmuskip}
\def\hp{\mskip0.5\thinmuskip}

\begin{document}

\title{Low-frequency measurement of the tunneling amplitude in a flux qubit}
\author{M.~Grajcar}
\altaffiliation[On leave from ]{Department of Solid State Physics,
Comenius University, SK-84248 Bratislava, Slovakia.}
\affiliation{%
Institute for Physical High Technology, P.O. Box 100239, D-07702
Jena, Germany}
\author{A.~Izmalkov}
\affiliation{%
Institute for Physical High Technology, P.O. Box 100239, D-07702
Jena, Germany}
\affiliation{%
Moscow Engineering Physics Institute (State University),
Kashirskoe sh.\ 31, 115409 Moscow, Russia}
\author{E.~Il'ichev}
\email{ilichev@ipht-jena.de}
\affiliation{%
Institute for Physical High Technology, P.O. Box 100239, D-07702
Jena, Germany}
\author{Th.~Wagner}
\affiliation{%
Institute for Physical High Technology, P.O. Box 100239, D-07702
Jena, Germany}
\author{N.~Oukhanski}
\affiliation{%
Institute for Physical High Technology, P.O. Box 100239, D-07702
Jena, Germany}
\author{U.~H\"ubner}
\affiliation{%
Institute for Physical High Technology, P.O. Box 100239, D-07702
Jena, Germany}
\author{T.~May}
\affiliation{%
Institute for Physical High Technology, P.O. Box 100239, D-07702
Jena, Germany}
\author{I.~Zhilyaev}
\altaffiliation[On leave from ]{Inst.\ of Microelectronic
Technology, Russian Academy of Science, 142432 Chernogolovka,
Russia.}
\affiliation{%
Institute for Physical High Technology, P.O. Box 100239, D-07702
Jena, Germany}
\author {H.E. Hoenig}
\affiliation{%
Institute for Physical High Technology, P.O. Box 100239, D-07702
Jena, Germany}
\author{Ya.S. Greenberg}
\altaffiliation[On leave from ]{Novosibirsk State Technical
University, 20~K.~Marx Ave., 630092 Novosibirsk, Russia.}
\author {V.I. Shnyrkov}
\altaffiliation[On leave from ]{B. Verkin Inst.\ for Low
Temperature Physics and Engineering, 310164 Kharkov, Ukraine.}
\affiliation{%
Friedrich Schiller University, Institute of Solid State Physics,
D-07743 Jena, Germany}
\author{D.~Born}
\affiliation{%
Institute for Physical High Technology, P.O. Box 100239, D-07702
Jena, Germany}
\author{W.~Krech}
\affiliation{%
Friedrich Schiller University, Institute of Solid State Physics,
D-07743 Jena, Germany}
\author{H.-G. Meyer}
\affiliation{%
Institute for Physical High Technology, P.O. Box 100239, D-07702
Jena, Germany}
\author{Alec Maassen van den Brink}
\affiliation{%
D-Wave Systems Inc., 320-1985 West Broadway, Vancouver, B.C., V6J
4Y3 Canada}
\author{M.H.S. Amin}
\affiliation{%
D-Wave Systems Inc., 320-1985 West Broadway, Vancouver, B.C., V6J
4Y3 Canada}

\date{\today}

\begin{abstract}
We have observed signatures of resonant tunneling in an  Al three-junction qubit, inductively coupled to a Nb $LC$ tank circuit. The resonant properties of the tank oscillator are sensitive to the effective susceptibility (or inductance) of the qubit, which changes drastically as its flux states pass through degeneracy. The tunneling amplitude is estimated from the data. We find good agreement with the theoretical
predictions in the regime of their validity.
\end{abstract}


\maketitle

Several groups, using different devices, have by now
established that superconductors can behave as macroscopic quantum
objects.\cite{macro,Wal00,Rabi} These are natural candidates for
a qubit, the building block of a quantum computer.

Qubits are effectively two-level systems with time-dependent
parameters. One of them is a superconducting loop with low
inductance $L$, including three Josephson junctions (a 3JJ
qubit).\cite{Mooij99} Its potential energy,
$U=\sum_{j=1}^{3}E_{\mathrm{J}j}(\phi_j)$, depends on the
Josephson phase differences~$\phi_j$ across the junctions.
Due to flux quantization $\sum_{j=1}^3\phi_j=2\pi\Phi_\mathrm{x}/\Phi_0$ (with $\Phi_\mathrm{x}$ the external magnetic flux and $\Phi_0=h/2e$ the flux quantum), only two $\phi_j$'s are independent.

For suitable parameters, $U(\phi_1,\phi_2)$ has two minima
corresponding to qubit states $\Psi^\mathrm{l}$ and
$\Psi^\mathrm{r}$, carrying opposite supercurrents around the
loop. These become degenerate for
$\Phi_\mathrm{x}=\frac{1}{2}\Phi_0$. The Coulomb energy $E_C$ ($\equiv e^2\!/2C$, with $C$ the capacitance of junction~1) introduces quantum uncertainty in the~$\phi_j$. Hence, near degeneracy the system can tunnel between the two potential minima. (Since $E_C \ll E_\mathrm{J}\equiv E_{\mathrm{J}1}$, we deal with a \emph{flux} qubit; $E_C\gg E_\mathrm{J}$ yields a \emph{charge} qubit. Coherent tunneling was demonstrated in both.)

\begin{figure}
\includegraphics[width=6cm]{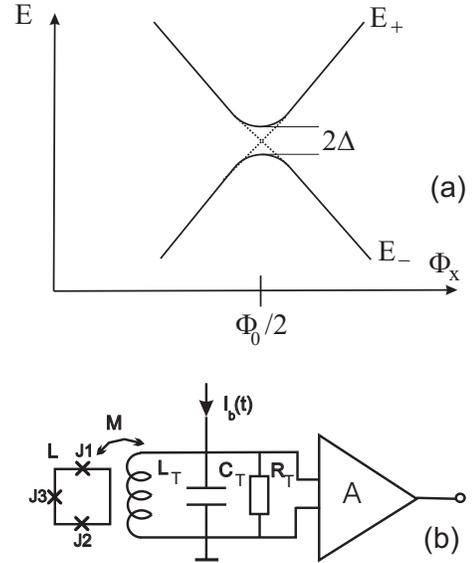}
\caption{(a)~Quantum energy levels of the qubit vs external flux.
The dashed lines represent the classical potential minima.
(b)~Phase qubit coupled to a tank circuit.}\label{fig:schem}
\end{figure}

In the basis $\{\Psi^\mathrm{l},\Psi^\mathrm{r}\}$ and near
$\Phi_\mathrm{x}=\frac{1}{2}\Phi_0$, the qubit can be described by
the Hamiltonian
\begin{equation}\label{eq01}
  H(t)= - \epsilon(t)\sigma_{\!z} -\Delta\sigma_{\!x}\;;
\end{equation}
$\Delta$ is the tunneling amplitude. At bias $\epsilon=0$ the two
lowest energy levels of the qubit anticross [Fig.~\ref{fig:schem}(a)], with a gap of~$2\Delta$. Increasing $\epsilon$ slowly enough, the qubit
can adiabatically transform from $\Psi^\mathrm{l}$ to
$\Psi^\mathrm{r}$, staying in the ground state~$E_-$. Since $dE_-/\hn d\Phi_\mathrm{x}$ is the persistent loop current, the
curvature $d^2\hn E_-/\hn d\Phi_\mathrm{x}^2$ is related to the
qubit's susceptibility. Hence, near degeneracy the latter will
have a peak, with a width given by $|\epsilon|\lesssim\Delta$.\cite{Greenberg02b} We present data demonstrating such behavior in an Al 3JJ qubit.

Our technique is similar to rf-SQUID readout.\cite{Silver67,Ilichev01} The qubit loop is inductively coupled to a parallel resonant tank circuit [Fig.~\ref{fig:schem}(b)]. The tank is fed a monochromatic rf signal at its resonant frequency~$\omega_\mathrm{T}$. Then both amplitude $v$ and phase shift $\chi$ (with respect to the bias current~$I_\mathrm{b}$) of the tank voltage will strongly depend on (A)~the shift in resonant frequency due to the change of the effective qubit inductance by the tank flux, and (B)~losses caused by field-induced transitions between the two qubit states. Thus, the tank both applies the probing field to the qubit, and detects its response.

The output signal depends on the tank's quality factor~$Q$. Using
superconducting coil, values as high as $Q\sim 10^3$ can be
obtained, leading to high readout sensitivity, e.g., in rf-SQUID
magnetometers.\cite{Danilov} Such a tank can therefore be used to
probe phase qubits.\cite{Greenberg02a} For small~$L$, the results
are summarized by\cite{Greenberg02b}
\begin{gather}\label{v}
  v=I_0\omega_\mathrm{T}L_\mathrm{T}Q/\sqrt{1+(2Q\xi)^2}\;,
    \displaybreak[0]\\
  \tan\chi=2Q\xi\;,\displaybreak[0]\\
  \xi(v,f_\mathrm{x})=\frac{k^2L}{2\Phi_0^2}\int_0^{2\pi}\!\frac{d\phi}{\pi}\,
    \cos^2\phi\,\frac{d^2E_-(f)}{df^2}\;,\label{xi-int}
    \displaybreak[0]\\
  f=f_\mathrm{x}
    +\frac{Mv}{\omega_\mathrm{T}L_\mathrm{T}\Phi_0}\sin\phi\;,\label{f-fx}  
\end{gather}
where $f_\mathrm{x}=\Phi_\mathrm{x}/\Phi_0-\frac{1}{2}$, $I_0$ is
the bias-current amplitude, and $k=M/\sqrt{LL_\mathrm{T}}$ is the
tank--qubit coupling coefficient, with $M$ ($L_\mathrm{T}$) the
mutual (tank) inductance. The ground-state curvature
is\cite{typo}
\begin{equation}
  \frac{d^2\hn E_-}{df^2}=-\frac{E_\mathrm{J}^2\Delta^2\lambda^2}
    {(E_\mathrm{J}^2\lambda^2f^2+\Delta^2)^{3/2}}\;,
    \label{Eq:dE}
\end{equation}
where $\lambda(\alpha,g)$ (with $g=E_\mathrm{J}/E_C$) is the
conversion factor in $\epsilon=E_\mathrm{J}\lambda
f$.\cite{lambda} If $I_0$ vanishes, $\xi=\half k^2\hn L\hp
d^2\hn E_-/\hn d\Phi_\mathrm{x}^2$ becomes an external parameter
accounting for the qubit susceptibility coupled to the tank. For
finite $I_0$, this has to be averaged over a bias cycle $0<\phi<2\pi$. The resulting integral (\ref{xi-int}) turns out to involve a weight $\cos^2\phi$, since the effective time-dependent coupling is $(k\dot{f})^2$ [cf.\ the $\phi$-derivative of Eq.~(\ref{f-fx})], proportional to the square of the voltage the tank induces in the qubit. The resulting equations are coupled and nonlinear, but readily solved numerically.

For the tank, we prepared a square-shape Nb pancake coil on an
oxidized Si substrate. The line width of the 20 windings was
2~$\mu$m, with a 2~$\mu$m spacing. Predefined alignment marks
allow placing a qubit in the center. For flexibility, only the
coil was made lithographically; an external capacitance
$C_\mathrm{T}$ is used to change $\omega_\mathrm{T}$ in the range
5--35~MHz. For the selected tank ($L_\mathrm{T}\approx50$~nH,
$C_\mathrm{T}\approx470$~pF), we obtained
$\omega_\mathrm{T}/2\pi=32.675$~MHz and $Q\approx725$ from the
voltage--frequency characteristic.

\begin{figure}
\includegraphics[width=8cm]{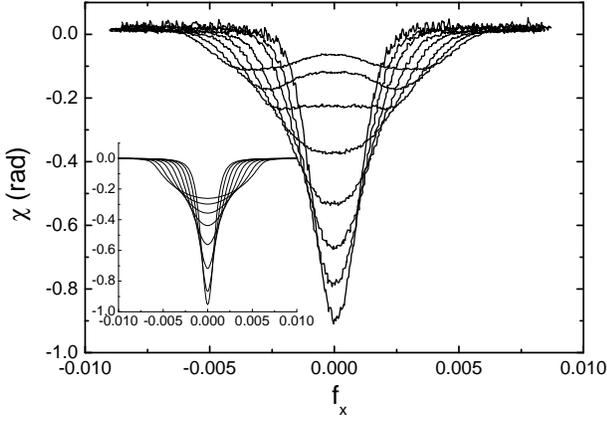}
\caption{Tank phase shift vs flux bias near degeneracy. From the
lower to the upper curve (at $f_\mathrm{x}=0$) the driving-voltage
amplitude $V_\mathrm{dr}\equiv I_0\omega_\mathrm{T}L_\mathrm{T}Q$ takes values 0.5, 1.0, 1.5, 1.9, 2.9, 3.5, 3.9~$\mu$V.
Inset: theoretical curves for $\Delta/h=650$~MHz, and $I_0=0.07$, 0.13, 0.20, 0.26, 0.39, 0.47, 0.53~nA.} \label{fig:Bias_dep}
\end{figure}

The 3JJ qubit structure was manufactured out of Al by conventional
shadow evaporation. The area of two of the junctions was estimated
using electron microscopy as $190\times650$~nm$^2$ while one is
smaller, so that $\alpha\equiv
E_{\mathrm{J}3}/E_{\mathrm{J}1,2}\approx0.8$. The critical current
was determined by measuring an rf-SQUID prepared on the same
chip\cite{Ilichev01} as
$I_\mathrm{c}=2eE_\mathrm{J}/\hbar\approx380$~nA. With
$E_C/h\approx3$~GHz, one finds  $g\approx60$ and
$\lambda\approx-4.4$. The loop area was 90~$\mu\mathrm{m}^2$, with
$L=38$~pH. We measured $v$ by a three-stage cryogenic amplifier,
placed at $\approx2$~K and based on commercial pseudomorphic high
electron mobility transistors. It was slightly modified from the
version in Ref.~\onlinecite{Oukhansky02} to decrease its back-action on
the qubit. The input-voltage noise was
$<0.6$~nV$\!/\sqrt{\mathrm{Hz}}$ in the range \mbox{1--35}~MHz.
The noise temperature was $\sim300$~mK at 32~MHz. The effective
qubit temperature due to the amplifier's back-action should be
considerably lower because of the small
$k\approx2\nobreak\cdot10^{-2}$.

\begin{figure}
  \includegraphics[width=8cm]{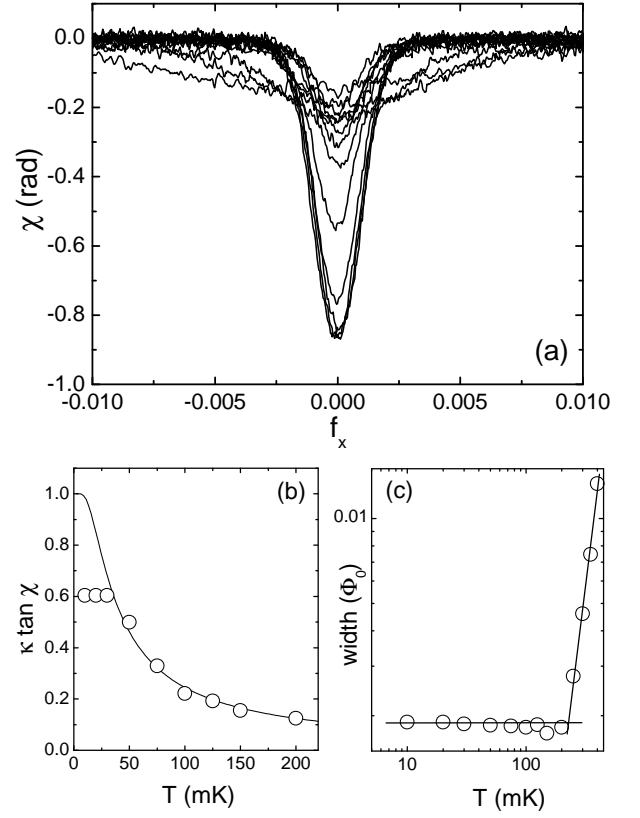}
  \caption{(a)~Tank phase shift vs flux bias near degeneracy and for $V_\mathrm{dr}=0.5~\mu$V. From the lower to the upper curve
  (at $f_\mathrm{x}=\nobreak0$) the temperature is 10, 20, 30, 50, 75,
  100, 125, 150, 200, 250, 300, 350, 400~mK.
  (b)~Normalized amplitude of $\tan\chi$ (circles) and $\tanh(\Delta/k_\mathrm{B}T)$ (line), for the $\Delta$ following from Fig.~\ref{fig:Bias_dep}; the overall scale $\kappa$ is a fitting parameter. The data indicate a saturation of the effective qubit temperature at 30~mK. (c)~Full dip width at half the maximum amplitude vs temperature. The horizontal line fits the low-$T$ ($<200$~mK) part to a constant; the sloped line represents the $T^3$ behavior observed empirically for higher~$T$.} \label{fig:Temp_dep}
\end{figure}

The $\chi(f_\mathrm{x})$ curves measured at various $I_0$ and a
mixing-chamber temperature $T=10$~mK are shown in
Fig.~\ref{fig:Bias_dep}. The narrow dip at $f_\mathrm{x}=0$
directly corresponds to the one in Eq.~(\ref{Eq:dE}), in line with the
qualitative picture below Eq.~(\ref{eq01}). With device parameters as
above, all quantities in Eqs.\ (\ref{v})--(\ref{Eq:dE}) are known, but
$\Delta$ only in principle: its exponential sensitivity to
$\alpha$ and especially $g$ makes it notoriously hard to calculate
\emph{a priori}. Hence, it is treated as a free parameter;
calculated curves for the best fit $\Delta/h=650$~MHz are shown in
the inset. For the largest $I_0$ the experimental and theoretical
curves disagree, for the rapid change of $\Phi_\mathrm{x}$ then
leads to Landau--Zener transitions\cite{LZ,andrei} suppressing the dip.

The $T$-dependence of $\chi$ is shown in Fig.~\ref{fig:Temp_dep}. For increasing $T$ the dip's amplitude decreases while, strikingly, its width is unchanged [Fig.~\ref{fig:Temp_dep}(c)]. Both are a simple manifestation of the Hamiltonian (\ref{eq01}) yielding $\langle\sigma_{\!z}\rangle=(\epsilon/\Omega)\*\tanh(\Omega/k_\mathrm{B}T)$,\cite{adia}
$\Omega=\sqrt{\epsilon^2+\Delta^2}$. This result of equilibrium
statistics of course assumes that the $t$\nobreakdash-dependence
of $\epsilon(t)$ is adiabatic. However, it \emph{does} remain valid if the full (Liouville) evolution operator of the qubit would contain standard Bloch-type relaxation and dephasing terms (which indeed are not probed\cite{Greenberg02b}) in addition to the Hamiltonian dynamics (\ref{eq01}), since the fluctuation--dissipation theorem guarantees that such terms do not affect equilibrium properties. Normalized dip amplitudes are shown vs $T$ in Fig.~\ref{fig:Temp_dep}(b) together with
$\tanh(\Delta/k_\mathrm{B}T)$, for $\Delta/h=650$~MHz
\emph{independently} obtained above from the low-$T$ width. The
good agreement strongly supports our interpretation, and is
consistent with $\Delta$ being $T$\nobreakdash-independent in the
relevant range.\cite{Han02} Of course, for higher $T$ the dip
will wash out; we observe a width $\propto T^3$ above a crossover
temperature $\approx 225$~mK. For $T$ of this order, deviations
from the two-state model can be expected, especially for
$f_\mathrm{x}\neq0$. This behavior outside the qubit regime has
not been pursued.

In conclusion, we have observed resonant tunneling in a
macroscopic superconducting system, containing an Al flux qubit
and a Nb tank circuit. The latter played dual control and readout
roles. The impedance readout technique allows direct
characterization of some of the qubit's quantum properties,
\emph{without} using spectroscopy.\cite{Wal00,Rabi} In a range
50$\sim$200~mK, the \emph{effective} qubit temperature
has been verified [Fig.~\ref{fig:Temp_dep}(b)] to be the same as the mixing chamber's (after $\Delta$ has been determined at low~$T$), which is often difficult to confirm independently.

MHSA and AMvdB are grateful to A.Yu.\ Smirnov and A.M. Zagoskin
for fruitful discussions, and to P.C.E. Stamp for the remark on
effective thermometry. MG wants to acknowledge partial support by the Slovak Grant Agency VEGA (Grant No.\ 1/9177/02).


\begin{thebibliography}{99}
\bibitem{typo} A minus sign is missing in Eq.~(17) of Ref.~\onlinecite{Greenberg02b}.
\end{thebibliography}
\end{document}